\documentclass[journal]{IEEEtran}

\usepackage{graphicx}
\usepackage{amsmath}
\usepackage{amssymb}
\usepackage{dsfont}
\usepackage{multicol}
\usepackage{svg}
\usepackage[utf8]{inputenc}
\usepackage{array}
\usepackage{wrapfig}
\usepackage{multirow}
\usepackage{tabularx}
\usepackage[mathlines,switch]{lineno}
\usepackage{etoolbox}
\usepackage{lipsum}
\usepackage{footnote}
\usepackage{pdflscape}
\usepackage{array}
\usepackage{makecell}
\usepackage{booktabs}
\usepackage{xcolor}
\usepackage{colortbl}

\definecolor{LULC}{RGB}{178,30,35}
\definecolor{LLLC}{RGB}{35,138,3}
\definecolor{RULC}{RGB}{8,26,183}
\definecolor{RMLC}{RGB}{160,13,165}
\definecolor{RLLC}{RGB}{12,143,148}

\ifCLASSOPTIONcompsoc
  \usepackage[nocompress]{cite}
\else
  \usepackage{cite}
\fi
\ifCLASSINFOpdf
\else
\fi
\makesavenoteenv{tabular}

\begin{document}
%
\title{Relational Modeling for Robust and Efficient Pulmonary Lobe Segmentation in CT Scans}
%
%
%
%

\author{Weiyi~Xie,~\IEEEmembership{}
        Colin~Jacobs,~\IEEEmembership{}
        Jean-Paul~Charbonnier,~\IEEEmembership{}
        Bram~van~Ginneken~\IEEEmembership{}
\IEEEcompsocitemizethanks{\IEEEcompsocthanksitem This work was supported by the Dutch Lung Foundation. We acknowledge the COPDGene  Study (ancillary study  ANC-398) for providing the data used. COPDGene is funded by Award Number U01 HL089897 and Award Number U01 HL089856 from the National Heart, Lung, and Blood Institute. The content is solely the responsibility of the authors. It does not necessarily represent the official views of the National Heart, Lung, and Blood Institute of the National Institutes of  Health. The COPD Foundation also supports the COPDGene project through contributions made to an Industry Advisory Board comprised of AstraZeneca, Boehringer Ingelheim, GlaxoSmithKline, Novartis, Pfizer, Siemens, and Sunovion. (Corresponding author: Weiyi Xie. Contacting email: Weiyi.Xie@radboudumc.nl). \protect
\IEEEcompsocthanksitem W.~Xie, C.~Jacobs, B.~van Ginneken are with the Diagnostic Image Analysis Group, Radboudumc, Nijmegen, the Netherlands.\protect
\IEEEcompsocthanksitem Jean-Paul  Charbonnier is with Thirona, Nijmegen, the Netherlands. \protect}

\thanks{}}

%
%

\markboth{}
{W.Xie \MakeLowercase{\textit{et al.}}: Object Relational Reasoning for Robust and Efficient Pulmonary Lobe Segmentation in CT Scans}
%



\IEEEtitleabstractindextext{%
\begin{abstract}
Pulmonary lobe segmentation in computed tomography scans is essential for regional assessment of pulmonary diseases. Recent works based on convolution neural networks have achieved good performance for this task. However, they are still limited in capturing structured relationships due to the nature of convolution. The shape of the pulmonary lobes affect each other and their borders relate to the appearance of other structures, such as vessels, airways, and the pleural wall. We argue that such structural relationships play a critical role in the accurate delineation of pulmonary lobes when the lungs are affected by diseases such as COVID-19 or COPD. 

In this paper, we propose a relational approach (RTSU-Net) that leverages structured relationships by introducing a novel non-local neural network module. The proposed module learns both visual and geometric relationships among all convolution features to produce self-attention weights.

With a limited amount of training data available from COVID-19 subjects, we initially train and validate RTSU-Net on a cohort of 5000 subjects from the COPDGene study (4000 for training and 1000 for evaluation). Using models pre-trained on COPDGene, we apply transfer learning to retrain and evaluate RTSU-Net on 470 COVID-19 suspects (370 for retraining and 100 for evaluation). Experimental results show that RTSU-Net outperforms three baselines and performs robustly on cases with severe lung infection due to COVID-19.
\end{abstract}

}

\maketitle
\IEEEdisplaynontitleabstractindextext
\IEEEpeerreviewmaketitle

\ifCLASSOPTIONcompsoc
\IEEEraisesectionheading{\section{Introduction}\label{sec:introduction}}
\else
\section{Introduction}
\fi
\IEEEPARstart{T}{he} human lungs consist of five disjoint pulmonary lobes. The right lung is composed of an upper, middle, and lower lobe, while the left lung only has an upper and a lower lobe. The lobes are separated by the pulmonary fissures, a double-fold of visceral pleura visible as a thin line on CT images. The lobes are functionally independent units because each has its own vascular and bronchial supply. As a result, the extent of the disease often varies substantially across lobes, and lobe-wise assessment of pulmonary disorders is of clinical importance. 

Computed Tomography (CT) is the best way to image the lungs in vivo. COVID-19, the pandemic disease caused by the SARS-Cov2 virus is straining healthcare systems worldwide.  A CT severity score can summarize the severity of the disease where each lobe is scored visually by radiologists on a scale from 0 to 5.  The summation of these scores quantifies lung involvement on a scale from 0 to 25 \cite{doi:10.1148/radiol.2020201473}. The score provides a tool to assess disease severity and progression, which further benefits clinical decision making. To automate the CT severity score, lobe segmentation in COVID-19 scans is needed. CT scans of COVID-19 patients are affected by extensive patchy ground-glass region and consolidations and may even show lobes or complete lungs filled with pleural fluid. Automated lobe segmentation is highly challenging in scans with such extensive pathological changes. 

Many automatic lobe segmentation approaches focused on finding visible fissures, assuming that the detected fissures equivalent to find the lobe segmentation by interpolation. Both early fissure enhancement filters \cite{kubo1999extraction,kubo2000extraction, wang2006pulmonary,pu2008computational} and more robust supervised learning methods \cite{van2007supervised} relied heavily on hand-crafted features, thus hard to generalize. Moreover, because incomplete fissures are very common \cite{raasch1982radiographic}, interpolation of boundaries based on visible fissures may not suffice to find the lobe borders reliably. Instead of finding fissures alone, anatomical relations between lobes and nearby airways, vessels, and the lung borders were exploited to account for incomplete fissures and damaged lung due to pathology \cite{kuhnigk2005new,van2010automatic,lassen2012automatic,bragman2017pulmonary}. 

Recent advances in convolution neural networks (CNN) provide a data-driven approach for more robust feature extraction in an end-to-end optimization process. Many works have successfully adopted CNNs in their lobe segmentation framework \cite{ferreira2018end,imran2019fast,wang2019automated,gerard2019pulmonary}. In \cite{ferreira2018end}, deep supervision was extensively used in the up-sampling path based on their V-Net design \cite{milletari2016v} along with the multi-tasking that segments lobe and lobe borders at the same time. \cite{imran2019fast} uses a relatively deeper architecture based on Dense-Net \cite{huang2017densely} to ensure a sufficient receptive field of extracted features. Global Geometric features were explored \cite{wang2019automated} as additional input channels to a convolution layer. 

The use of multi-resolution input in a two-stage cascading CNNs to extract both global and local features has been proposed for lobe segmentation in CT by Gerard et al.~\cite{gerard2019pulmonary}. Their first stage network was trained on low-resolution images to learn global features from the entire scan. The global features were added to the second stage network to provide contextual guidance, while the second stage network was designed to focus on capturing local details at a high resolution. Their framework has also been successfully applied for pulmonary fissure and lung segmentation tasks \cite{gerard2018fissurenet,gerard2020multi}. In this work, we also employ a two-stage approach, that is trained in an end-to-end fashion.

Although existing CNN approaches have achieved superior performance in lobe segmentation comparing to the feature engineering approaches, they may still be inefficient and limited in relational reasoning, such as capturing the interlobar relationships and other long-range relationships between lobes and other structures in the CT image. CNN approaches assumed that such relationships between objects and object parts in semantic segmentation could be implicitly learned directly from the CNN training process. 

However, as \cite{hu2019local,zhang2019dynamic} have pointed out, the hierarchical feature representation computed using a sequence of stacked convolutional layers can be highly inefficient in inferring relations between convolution features. As higher-level features in CNNs commonly represent objects and object parts, instead of aggregating these features based on their semantic interactions, convolutional filters act as templates, where features are aggregated depending on the filter weights. This may cause inefficiency in capturing relations between features because filters weights are not invariant to permutations of features. In addition, convolution filters are limited to capture long-range relations due to the use of local kernels. 

CT findings in patients of a COVID-19 infection \cite{bernheim2020chest,shi2020radiological} often include multiple regions with focal pathological changes, ranging from ground-glass to consolidations to organizing pneumonia. These changes occur more often in the lower lobes. Here the lobar boundaries can be deformed substantially. In these cases, information from other regions in the CT image may be crucial for locating and delineating a target lobe. Therefore, in this paper, we introduce a novel non-local neural network module to model the global structured relationships for pulmonary lobe segmentation. The proposed non-local neural network module computes a feature response at one location using both appearance and geometric features from all other positions at the scan-level. We call this approach a Relational two-stage U-net, or RTSU-Net, for short.

The main contributions of this paper are as follows:
\begin{itemize}\item We propose a novel non-local neural network module that can capture the global structured relationships between object and object parts in terms of their visual and geometric features for the lobe segmentation. The proposed RTSU-Net is robust and produces accurate lobe segmentations even for scans with severe pathology.
\item We used a multi-resolution framework similar to \cite{gerard2018fissurenet,gerard2019pulmonary,gerard2020multi}, however, we train both stages in an end-to-end fashion. This gives the RTSU-Net the ability to capture the global object relationships at the full scan level from the first stage network while extracting local details at the second stage simultaneously in the same optimization process. 
\item RTSU-Net is fast and memory-efficient, considering it consists of a cascade of two CNNs. RTSU-Net requires only a standard GPU with 12GB memory to train and takes around 30 seconds to produce lung and lobe segmentations for a full thoracic CT scan at test time. The time consumption includes the CNN inference time, pre-processing, and post-processing, excluding the time spent on IO. 
\end{itemize}

\subsection{Related Work}
Although convolutional neural networks (CNNs) have achieved superior performance in a wide range of medical imaging segmentation tasks \cite{cciccek20163d,milletari2016v}, they are still limited in modeling object relationships, especially the long-range interactions. Several techniques have been proposed to account for the missing capability of relational reasoning in CNNs. 

Poudel et al.~\cite{poudel2016recurrent} introduced recurrent neural networks to aggregate features across the axial slices for cardiac segmentation in multi-slice MRI. A known issue with recurrent network networks is that they suffer from vanishing gradients \cite{pascanu2013difficulty} and therefore are hard to train. The object relations could also be explicitly defined using Graph Models such as dense conditional random fields (CRF) \cite{kamnitsas2017efficient}. However, due to their heavy computational demands, dense CRFs are often only used as the post-processing steps and optimized separately on a heuristic basis, making it hard for this approach to scale well.

Attention is widely used for various tasks such as machine translation, image and video classification, object detection, and semantic segmentation. Self-attention methods \cite{bahdanau2014neural,vaswani2017attention} capture contextual dependencies between words by computing the embedding at one word by a weighted summation of embeddings at all words in sentences. As one of the self-attention applications, a non-local neural network was proposed for semantic segmentation \cite{wang2018non} by computing a global self-attention map for each feature based on all the other features in an input CNN feature map. The attention weights were determined by predefined similarity measurements between pairwise features in a linear-projected subspace, as an efficient way of modeling their conceptual relationships. 

There are several recent extensions of this non-local method in semantic segmentation. CCNet \cite{huang2018ccnet} was proposed to employ a simple criss-cross trick, which reduces the space and time complexity of the non-local module from $O((H\times W)\times(H\times W))$ to $O((H\times W)\times(H+W-1))$ in two-dimensional images. Hu et al.~\cite{hu2019local} aggregated features based on both visual and geometric correspondence in a locally connected aggregation graph, thus lacking long-range relationships. A dynamical aggregation graph was proposed in \cite{zhang2019dynamic} to capture both short and long-range relationships, but no geometric correspondence between features was used. 

Our approach is motivated by the above works. Our self-attention module uses the criss-cross trick to collect global structured relationships between object and object parts in terms of their visual and geometric correspondence in the feature representation.

\section{Data}\label{data}

\begin{table}
  \caption{Characteristics of the two data sets used in this study. (a) lists the distribution of GOLD stages and other classes, see \cite{regan2011COPDGene} in the COPD data set. (b) gives the distribution of CO-RADS scores \cite{doi:10.1148/radiol.2020201473} across the training and test sets. CO-RADS score 1-6 indicates the level of suspicion for COVID-19 positive disease, ranging from very low, low, equivocal, high, very high, and confirmed positive from the reverse-transcription polymerase chain reaction (RT-PCR) tests, respectively.}
  \centering
  \begin{tabular}{lll}
    \multicolumn{3}{c}{(a) COPD set GOLD stages}\\
    \toprule
    GOLD stages     & \#subjects for training     & \#subjects for testing \\
    \midrule
    GOLD0     & 1709 &  433     \\
    GOLD1     & 319 &  80     \\
    GOLD2     & 734 &  184     \\
    GOLD3     & 441 &  110     \\
    GOLD4     & 226 &  57     \\
    Non Spirometry  &30 & 2 \\
    Non Smoking     & 45 &  11     \\
    PRISm     & 496 &  123     \\
    \bottomrule
     Total     & 4000 &  1000     \\
    \bottomrule
&&\\
  \multicolumn{3}{c}{(b) COVID-19 set CO-RADS}\\
  \toprule
  CORADS     & \#subjects for training 
  & \#subjects for testing \\
  \midrule
  1     & 158 &  23     \\
  2     & 46 &  9     \\
  3     & 47 &  20     \\
  4     & 30 &  16     \\
  5     & 65 &  24     \\
  6    & 24 &  8     \\
  \bottomrule
Total & 370 & 100\\
  \bottomrule
\end{tabular}
 
 \label{tab:datameta}
\end{table}
CT scans used in this study were obtained from two sources. We refer to the first set as the COPD set and the second set as the COVID-19 set.

A large set of scans from subjects with COPD, ranging from mild to very severe, was obtained from the COPDGene study \cite{regan2011COPDGene}. This is a clinical trial with data from 21 imaging centers in the United States. In total, COPDGene enrolled 10,000 subjects. Each subject underwent both inspiration and expiration chest CT. Image reconstruction uses sub-millimeter slice thickness and in-plane resolution, with edge-enhancing and smooth algorithms. Data from COPDGene is publicly available and can be retrieved after submitting an ancillary study proposal (ANC-398 was used for this work).

We randomly selected 5000 subjects and used only Phase I inspiration CT scans (one scan per subject). Subjects were randomly grouped into a training set ($n$ = 4000) and a test set ($n$ = 1000). Slice thickness ranged from 0.625-0.9mm and pixel spacing from 0.478-1.0mm. Most scans were performed using 200mAs a tube voltage of 120kVp and B31f and B35f reconstruction kernels. The CT protocols are detailed in \cite{regan2011COPDGene}.   

The other data set was obtained from Radboud University Medical Center, Nijmegen, the Netherlands. On March 18, 2020, this institution implemented a low-dose non-contrast CT protocol and all patients who arrived at the hospital with suspicion of COVID-19 disease and inpatients for whom COVID-19 was considered a possibility underwent CT. In accordance with local guidelines, we only included scans from subjects who did not object to the use of their scans for research purposes and we worked with anonymized data. Permission for research use was obtained from our review board (file number CMO 2016-3045, Project 20027). It is the intention to share these scans via a national Dutch COVID-19 database. 

We randomly selected 470 subjects and used one scan per subject by selecting the CT scan of the smallest slice thickness in a study. Scans have a pixel spacing between 0.5mm to 0.9mm and a slice thickness of 0.5 mm. Scans were performed using X-ray tube current ranging from 10mA to 493mA and a tube voltage of either 100 or 120kVp. Convolution kernels in reconstruction were lung kernels (FC83, FC86). 370 of these scans were used for training and the other 100 for testing. 

See Table \ref{tab:datameta} for the distribution of GOLD stages in the training and the test set for the COPD set and the distribution of CO-RADS scores \cite{doi:10.1148/radiol.2020201473} from the COVID-19 set. The CO-RADS scores defined the level of suspicion COVID-19 and were extracted from the radiology reports. Complete individual results of reverse-transcription polymerase chain reaction (RT-PCR) tests were not available at the time of anonymization of the data, but it is known that the majority of the test cases were positive for COVID-19 (these test cases overlap with the data used in \cite{doi:10.1148/radiol.2020201473}). 

From the two training data sets, we selected 100 scans as the validation set for the COPD set, and we selected 50 scans for validation from the COVID-19 set for retraining all the models. 


\subsection{Reference Standard}
Lobe segmentation references were obtained from Thirona, a company that specializes in chest CT analysis. First, automated segmentation of the left and right lung was generated using a commercialized software (LungQ, Thirona, Nijmegen, NL), followed by manual refinement if needed. Second, automatic algorithms \cite{van2007supervised,van2009automatic,van2010automatic} were used to extract the lobar boundary with possible interpolation for incomplete fissures using information from nearby airways and vessels. Next, the automatically found lobar boundaries were manually corrected separately for the left and the right lung, by trained analysts with at least one year experience in annotating pulmonary structures on CT. Analysts repeatedly adjusted the control points on the auto-generated lobar boundaries until the updated lobar boundaries were satisfactory. All analysts have a medical background and have received extensive training in lung anatomy and segmenting lobes in CT imaging. In case of doubt, radiologists could be consulted.

\section{Methods}
We define the lobe segmentation problem as a voxel-wise classification problem. Given a scan $I$, the goal is to predict the voxel label $\hat{l_{i}}$ for every spatial location $i$, where $\hat{l_{i}} \in$ the label set $L= \{0, 1, 2, 3, 4, 5\}$ representing the background, left upper, left lower, right upper, right lower, and the right middle lobe, respectively.

\begin{figure*}
\centering
\includegraphics[width=0.85\textwidth]{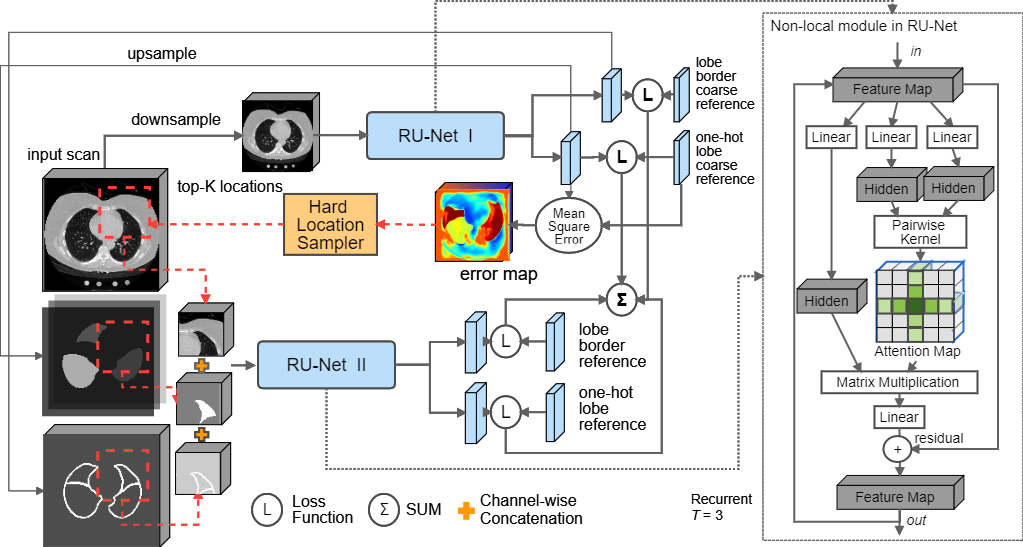}
\caption{The overview of our lobe segmentation framework with a cascade of two CNNs. At each stage, a CNN (RU-Net) uses the proposed non-local module to capture the structured relationships between objects and object parts. The output from the RU-Net I is concatenated with the cropped 3D patches as the input for RU-Net II.} 
\label{fig:framework}
\end{figure*}
In this paper, we use a multi-resolution approach with two cascaded CNNs to capture both global context and local details for the lobe segmentation, as proposed in \cite{gerard2019pulmonary}. Our framework is depicted in Fig.~\ref{fig:framework}. 

Besides the use of a multi-resolution framework, we introduce a novel non-local module to capture structured relationships and our efficient network design allows end-to-end training of our multi-resolution framework. For each CNN, we place our proposed non-local module to aggregate relational information for the features at the coarsest resolution as these features commonly represent high-level semantics such as objects and objects parts \cite{girshick2014rich}. The proposed non-local module computes visual and geometric correspondence between these features, naturally modeling relationships between objects and object parts. The use of geometric information is inspired by \cite{wang2019automated}. Also, the proposed non-local module can enlarge the receptive fields of these features because the computation of one non-local response involves all features in the feature map. We refer the CNN with the proposed non-local module at each stage to as relational U-Net (RU-Net), and the details are explained later in this section. 

\subsection{Cascading relational U-Nets} The first RU-Net reads an input scan at a down-sampled resolution to coarsely segment the lobes and lobe borders. These coarse outputs are subsequently upsampled to a higher resolution by trilinear interpolation. The high-resolution input scan and the output of the first RU-Net are concatenated and cropped into 3D patches as the input to train the second RU-Net to precisely segment lobes and lobe borders. The cascade of two relational U-Nets is trained end-to-end, allowing both local details and scan-level context to be learned in the same optimization process. Furthermore, we use the errors found in the predictions of the first RU-Net to optimally sample 3D patches for training the second stage, which encourages the second RU-Net to focus on the regions where the first RU-Net fails. This technique can be seen as a form of online hard example mining. \cite{shrivastava2016training}.

\subsection{Relational U-Net}\label{RU-Net}
The relational U-Net architecture (RU-Net) is a 3D U-Net architecture \cite{cciccek20163d} with a smaller number of convolution filters and an additional non-local module. The RU-Net has three down-sampling layers in the encoding path, and each layer consists of two convolutions and a max-pooling operation. Following the down-sampling path, two more convolutions are used to double the number of convolution filters. We then place the non-local module before up-sampling. In the up-sampling path, three layers are used to reconstruct the resolution, and each contains one tri-linear interpolation, followed by two convolutions to reduce the interpolation artifacts. In the end, features are reshaped via a single 1x1x1 convolution in two parallel output branches, and each corresponds to a different learning objective; one produces 6-channel softmax probabilities for segmenting the background and the five lobes. The other provides a single channel probability map by sigmoid function for predicting the lobe border. Features from 3x3x3 convolutions are batch normalized and activated via a rectifier linear unit (ReLU). No dropout is used.

The first RU-Net uses padded convolutions, whereas the second uses valid convolutions. The details regarding RU-Net network architecture on both stages are provided in Table \ref{tab:RUNetParam}, where the names of the down-sampling layers are prefixed with 'Down', and the name of up-sampling layers are prefixed with 'Up'. The numbers listed are based on the execution order.
\begin{table}
  \caption{Architectures for the first and the second stage of the relational U-Nets. The convolution filters are named by the kernel sizes $K$ and number of filters $N$ as $K\times K\times K,N$ (stride 1 for all). Non-local linear embedding parameters are defined in Eqs.~(\ref{non-local-geo-ref}) and (\ref{ccnet-geo-recurrent}). $\parallel$ denotes the operation performed in dual paths.}
  \centering  
  \vspace*{2mm}
  \begin{tabular}{m{4em} |  m{8em} |m{8em} }
    \hline
    Layer  &   RU-Net I & RU-Net II\\
    \hline
    \multirow{2}{4em}{Down1}   & 3x3x3,1-16 3x3x3,16-24 & 3x3x3,8-24 3x3x3,24-48\\\cline{2-3}
      &   \multicolumn{2}{c}{2x2x2 max pool, stride 2}\\
    \hline
    \multirow{2}{4em}{Down2}  & 3x3x3,24-24 3x3x3,24-48 & 3x3x3,48-48 3x3x3,48-96\\\cline{2-3}
      &  \multicolumn{2}{c}{2x2x2 max pool, stride 2}\\
    \hline
    \multirow{2}{4em}{Down3} & 3x3x3,48-64 3x3x3,64-128 & 3x3x3,96-96 3x3x3,96-192\\\cline{2-3}
      &  \multicolumn{2}{c}{2x2x2 max pool, stride 2}\\
    \hline  
    Bridge   & 3x3x3,128-128 3x3x3,128-256 & 3x3x3,192-192 3x3x3,192-384\\
    \hline 
    Non-local  &    $W_{\theta}$,$W_{\phi}$$\in$$\mathbb{R}^{256\times 32}$  $W_{\omega}$,$W_{\rho}$$\in$$\mathbb{R}^{3\times 32}$ $W_{r}$$\in$$\mathbb{R}^{32\times 256}$& $W_{\theta}$,$W_{\phi}$$\in$$\mathbb{R}^{384\times 32}$ $W_{\omega}$,$W_{\rho}$$\in$$\mathbb{R}^{3\times 32}$ $W_{r}$$\in$$\mathbb{R}^{32\times 384}$\\
    \hline
    \multirow{2}{4em}{Up1}  & 3x3x3,384-128 3x3x3,128-128 & 3x3x3,576-192 3x3x3,192-192\\\cline{2-3}
      &  \multicolumn{2}{c}{trilinear interpolation x2}\\
    \hline
    \multirow{2}{4em}{Up2}  & 3x3x3,176-48 3x3x3,48-48 & 3x3x3,288-96 3x3x3,96-96\\\cline{2-3}
      & \multicolumn{2}{c}{trilinear interpolation x2}\\
    \hline  
    \multirow{2}{4em}{Up3} & 3x3x3,72-24 3x3x3,24-24 & 3x3x3,144-48 3x3x3,48-48\\\cline{2-3}
      & \multicolumn{2}{c}{trilinear interpolation x2}\\
    \hline 
    Output   & 1x1x1,6 $\parallel$ 1x1x1,1 & 1x1x1,6 $\parallel$ 1x1x1,1\\
    \hline
    MAC & 5.71 G & 8.79 G \\
    \hline
    \#Parameter & 3.85M & 9.24 M \\
    \bottomrule
  \end{tabular} 
\label{tab:RUNetParam}
\end{table}


\subsection{The Non-Local module}
The original non-local neural network \cite{wang2018non} for semantic segmentation computes the feature response at a position as a weighted sum of the features at all locations in the input feature maps as 
\begin{equation}\label{non-local}
    y_{i} = \frac{1}{\zeta(x)}\sum_{\forall j}f(x_{i}, x_{j})g(x_{j}),
\end{equation}
where $y_{i}$ at location $i$ is computed as a weighted sum using the correspondence between the feature $x_{i}$ at the location $i$ and all features indexed by $j$ in the input feature map $x$. The feature correspondence between feature $x_{i}$ and $x_{j}$ is also called the self-attention in this context, computed by the pairwise function $f$, which is used to weigh the feature embedding $g(x_{j})$ before normalizing by $\zeta(x)$. For simplicity, $g$ is set to a linear projection: $g(x_{j})=W_{g}x_{j}$, and the pairwise function $f$ can be the embedded Gaussian function using linear embeddings defined as $f(x_{i}, x_{j})=e^{(W_{\theta}x_{i})^T(W_{\phi}x_{j})}$. We set the normalizing factor as $\zeta(x)=\sum_{\forall j}f(x_{i}, x_{j})$. Then $y$ becomes the softmax computation along the dimension $j$ written, in matrix multiplication form, as $y=softmax(x^TW^T_{\theta}W_{\phi}x)g(x)$. To make the input and output of the non-local module the same size, the $y_{i}$ is reshaped to have the same dimensions as the input $x_{i}$ by applying the linear reconstruction function $r, r(y_{i})= W_{r}y_{i}$. Therefore, the non-local response at location $j$ can be written as $z_{i}=W_{r}y_{i}+x_{i}$.

The feature response $z_{i}$ automatically achieves a global receptive field with respect to the input. The computed self-attention map $f(x_{i}, x_{j})$ captures the feature correlations, as relevant features would have higher attention responses.

However, the original non-local module disregards the geometric correspondence between features, while \cite{wang2019automated} shows that introducing geometric coordinates improves the performance on lobe segmentation. Hence, we propose to compute non-local responses with a geometric term. Here, we denote $\mu_{i}, \mu_{j}$ as geometric coordinates for the position $i$ and $j$. $\mu_{i}$ is the center coordinate of the receptive field of the feature at position $i$ with respect to the original input image and rescaled to $\left[ 0\sim1\right]$ range by the size of the original input image. We note that if the feature map is produced from a cropped input, the center coordinate of the receptive field is then shifted according to the 3D patch offset to the original input image. The rescaled geometric coordinates are then shifted by 0.5 to have zero mean. $\tau(\mu_{i}, \mu_{j})$ is the pairwise function for measuring correlations. Then, the non-local response with geometric terms is defined as: 
\begin{equation}\label{non-local-geo}
    y_{i} = \frac{1}{\zeta(x, \mu)}\sum_{\forall j}(f(x_{i}, x_{j}) + \tau(\mu_{i}, \mu_{j}))g(x_{j}),
\end{equation}
A similar reparameterization can be applied using the softmax function row-wise under linear projections to reformulate Equation \ref{non-local-geo} into matrix multiplications:
\begin{equation}\label{non-local-geo-ref}
    y = softmax(x^TW^T_{\theta}W_{\phi}x + max(0, \mu^TW^T_{\omega}W_{\rho}\mu))g(x),
\end{equation}
where $f(x_{i}, x_{j})$ is parameterized as a dot product in a subspace projected using the linear transformation matrix $W_{\theta}$ and $W_{\phi}$. Similarly, $W_{\omega}$ and $W_{\rho}$ are linear transformations that are used to project the geometric features $\mu$ into a subspace where their correspondence is measured by the pairwise kernel function $\tau$, $\tau(\mu_{i},\mu_{j})=max(0, \mu^TW^T_{\omega}W_{\rho}\mu)$. Such correspondence is then trimmed at 0, to restrict geometric relations within a certain threshold. 

The Equation \ref{non-local-geo-ref} however, has high computational cost because the self-attention map requires computing $x^TW^T_{\theta}W_{\phi}x$ and $\mu^TW^T_{\omega}W_{\rho}\mu$ on all pairs of locations. Each term has complexity in time and space of $O(C\times W^{2}\times H^{2}\times D^{2})$ where $C$ is the dimension of linear projected subspace and $W, H, D$ denotes the width, height, and depth of a 3D feature map. To reduce computational complexity, we adopt the criss-cross trick \cite{huang2018ccnet}, which has a time and space complexity of $O((C\times W\times H\times D)\times (H+W+D-2))$. In CCNet, Equation \ref{non-local-geo} is modified to:
\begin{equation}\label{ccnet-geo}
    y_{i} = \frac{1}{\zeta(x, \mu)}\sum_{j\in\Omega{j} }(f(x_{i}, x_{j}) + \tau(\mu_{i}, \mu_{j}))g(x_{j}),
\end{equation}
where $\Omega{j}$ indicates the neighboring voxels with respect to $j$ under criss-cross connectivity, such sparse connectivity requires having three recurrent criss-cross modules to cover all spatial locations in computation.

Given the input feature $x_{i}$, the non-local response $z_{i}^{t}$ for a feature location $i$ at each $t$-th recurrent criss-cross module can be written as follows:
\begin{equation}\label{ccnet-geo-recurrent}
\begin{array}{l}
z_{i}^{t} =  
\begin{cases}
    x_{i} & \text{if } t=0\\
    W_{r}y_{i}^{t-1} + z_{i}^{t-1} &  \text{if } t=1,2,\ldots,T\\
\end{cases} \\
y_{i}^{t} = \frac{1}{\zeta(z^{t}, \mu)}\sum_{j\in\Omega{j} }(f(z_{i}^{t}, z_{j}^{t}) + \tau(\mu_{i}, \mu_{j}))g(z_{j}^{t})
\end{array}
\end{equation}
At each recurrent step, the non-local response $z_{i}^{t}$ is used as the input feature for computing the non-local response for the next recurrent step. For the size of scans used in this work, full global context can be achieved with three recurrent steps for a 3D input feature map. Therefore, we set $T=3$.

\subsection{Online Hard Example Mining}
As shown in Fig.~\ref{fig:framework} using the red dashed lines, we compute the mean square errors (MSE) between the lobe-wise softmax probabilities of the first RU-Net and the lobe reference standard. We then go through all sliding window 3D patches, and find $K$ patches with the highest integral of MSE and use them for training the second RU-Net. 

$K$ is set to 1.0 such that all patches are used to train at the beginning and continuously reduced until it reaches a coverage of only approximately 20\% of the scan volume at the end of the training process. The proposed online hard example mining does not introduce extra forward and backward passes on the network, therefore the additional computational cost is trivial.

\subsection{Learning Objectives}
There are two learning objectives for each RU-Net: lobe segmentation and lobe border segmentation, inspired by \cite{ferreira2018end,gerard2019pulmonary}. Therefore, the final loss function is a summation of four terms, and each is the generalized Dice loss\cite{sudre2017generalised}. The lobe border reference is pre-computed from the lobe reference by detecting object boundaries.

Let $r$ be the segmentation reference with $n$-th voxel values $r_{ln}$ for the class label $l$ and $\hat{r}_{ln}$ be the predicted probabilistic map for the label $l$ over $n$-th image voxel, then the generalized Dice loss is defined as:
\begin{equation*}
    GLD = 1 - 2\frac{\sum_{l}w_{l}\sum_{n}r_{ln}\hat{r}_{ln}}{\sum_{l}w_{l}\sum_{n}r_{ln} + \hat{r}_{ln}},
\end{equation*}
with $w_{l} = 1/(\sum_{n}^{N_{l}}r_{ln})^2$, where $N_{l}$ the in total number of voxels for the class label $l$ in the segmentation reference. $w_{l}$ is to re-balance learning against the variance in object volumes.

\section{Experiments}
As the COVID-19 pandemic emerged only recently, it was not possible to obtain a large amount of CT scans with annotations of COVID-19 patients. Therefore, we used a transfer learning approach in our experiments. For training of the models on the COVID-19 data, the models were initialized with the trained weights from our models developed on the COPD data set.

\subsection{Training details}\label{subsec:traincopdgene}


Training, validation, and testing of each experiment were carried out on a machine with a NVidia TitanX GPU with 12 GB memory. The methods were implemented using Python 3.6, Pytorch 1.1.0 library \cite{paszke2017automatic}. The trainable parameters of each method were initialized using Kaiming He initialization when training from scratch \cite{he2015delving} and were optimized using stochastic gradient descent with a momentum of 0.9, and the initial learning rate set to 10e-6.
The initial models were trained using CT scans from the COPD data set. Therefore, these models may not be familiar with the visual patterns in COVID-19 scans. For efficiently training on new visual patterns, all models were retrained using a combined loss between the generalized Dice loss (as we used to train the initial models) and top-$K$ cross-entropy loss where $K$ is set to 30\% of all voxels in the input. The top-$K$ cross-entropy loss was implemented simply as the voxel-wise cross-entropy loss but selecting only $K$ voxels with the largest cross-entropy to back-propagate.   
\subsection{Comparison with previous work}
We compared our approach with three baselines, the well-known 3D U-Net and two recently published methods for lobe segmentation in CT. 
.
\subsubsection{3D U-Net}\label{cicek3d}
We implemented 3D U-Net following the original paper \cite{cciccek20163d}. The input is a mini-batch of two $132\times132\times132$ 3D patches randomly cropped from the pre-processed scan (refer to \ref{subsec:preandpost}). As a result of using valid convolutions, the output of this network is $44\times 44 \times 44$ voxels. During test time, the softmax probabilities of all 3D patches are tiled together by sliding over the entire scan without overlaps to build up a scan-level probability map. The final prediction is then made by assigning each voxel to the label with the highest probability. 
\subsubsection{FRV-Net and PDV-Net} 
We compare the proposed method with two existing end-to-end lobe segmentation methods. FRV-Net \cite{ferreira2018end} follows the design of the V-Net \cite{milletari2016v} and extensively uses the idea of deep supervision at almost all scales in the up-sampling pathways. PDV-Net \cite{imran2019fast} uses dense connections, following the DenseNet \cite{huang2017densely}, to design their network with a considerably large receptive field to capture contextual information. PDV-Net takes the entire CT scan as the input, thus potentially capable of learning the global information. Note that these two works have specific pre-processing and post-processing strategies. The input scan in FRV-Net is resized into a fixed size of $128\times256\times256$ and intensities are clipped into the range $\left[ -1000\sim400\right]$ HU. In PDV-Net, the input scan is resized into 128 $\times$ 512 $\times$ 512. We implemented both architectures following the paper at our best efforts.  

\subsection{Ablation studies}
To assess the contribution of the proposed non-local module in RTSU-Net, we performed several ablation studies. During these experiments, the models were trained from scratch using the COPD training set and retrained on COVID-19 and performance is measured on the COPD test set of 1000 cases and 100 COVID-19 cases. 
The performance of our proposed model was assessed  without the geometric features in the non-local module, and without the non-local module in the relational two stage U-Net framework.
\subsection{Pre-processing and post-processing}\label{subsec:preandpost}
All training and test scans were standardized by clamping intensity values to the $\left[ -1200\sim400\right]$ range before re-scaling into $\left[ 0\sim 1\right]$. Then all scans were down-sampled using trilinear interpolation to have a $256\times 256$ in-plane resolution while z-spacing is adjusted to make the scan isotropic. 

The input size of the second CNN for our proposed method consisted of two $116\times116 \times116$ sized 3D patches. The pre-processed scan was down-sampled by a factor of 2 using trilinear interpolation as the input for the first stage (padding with zeros are needed if the size on z axial is not divisible by 16). The softmax probability outputs of all 3D patches at the second stage were tiled together by sliding over the entire scan without overlaps to produce a scan-level probability map, which is used to generate the final prediction by assigning each voxel to the label with the highest probability.

As a post-processing step, the predictions were then up-sampled by nearest neighbor interpolation to match the original resolution of the scans. All evaluations are performed by using predictions and reference segmentations at the original resolution.
\subsection{Evaluation Metrics}
The Intersection over Union (IOU), and average symmetric surface distance (ASSD) between predictions and segmentation references were used for quantitative evaluation of segmentation performance. The IOU between two binary masks $X, Y$ is defined as:
\begin{equation*}
IOU(X, Y) = \frac{|X\cap Y|}{|X\cup Y|},
\end{equation*}
Denote two surfaces as $S_{X}$,$S_{Y}$ from the masks $X, Y$, and coordinate indices on the surface as $x$,$y$. The average symmetric surface distance (ASSD) is defined as:
\begin{equation*}
\begin{array}{l}
ASSD(X,Y) = 
\frac{\sum_{x\in S_{X}}\min_{y\in S_{Y}}d(x, y) + \sum_{y\in S_{Y}}\min_{x\in S_{X}}d(y, x)}
    {|S_{X}| + |S_{Y}|}
\end{array}
\end{equation*}
with $d(\cdot)$ being the Euclidean distance, and $|S_{X}|$ and $|S_{Y}|$ the surface area for $S_{X}$ and $S_{Y}$, respectively. 

Besides the lobe-based measurements, we also evaluated the performance of all models in the lung segmentation task by taking the union of all lobes as the lung. 
Furthermore, we add a metric to measure fissure alignment by computing the average symmetric surface distance in the interlobar borders between the predictions and the segmentation references.

The overall performance of the method was evaluated by computing the average of the per-lobe metrics. A Wilcoxon signed-rank test was employed to assess whether the performance difference was statistically significant ($p < 0.01$ with Bonferroni correction).

Also, we computed the number of Multi-Adds operations (MAC) and the number of parameters to assess computational efficiency. We also provide a comparison with independent human readers on a subset of 100 subjects from the COPD data only. 

 \begin{figure}
     \includegraphics[width=\linewidth]{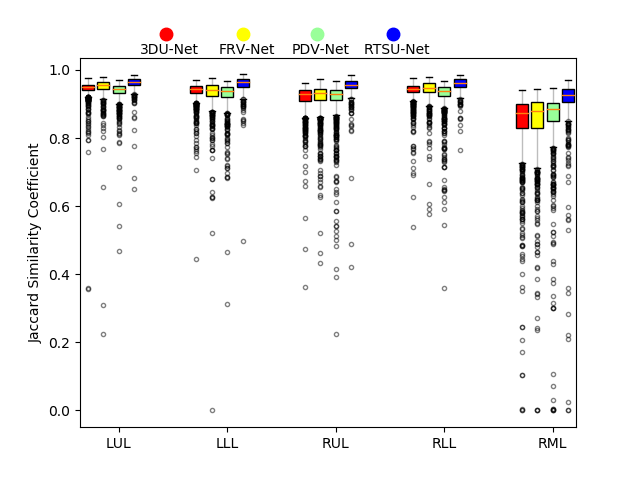}
     \includegraphics[width=\linewidth]{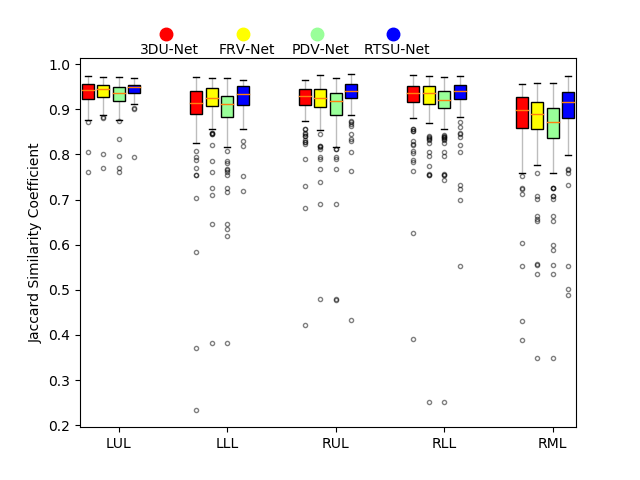}
     \caption{Box and whisker plots of IOU per-lobe for different methods on the COPD data set (top) and the COVID-19 data set (bottom).}
     \label{fig:boxjaccard}
\end{figure}
 
\section{Results}
\subsection{Quantitative results}
\begin{table*}[t]
\caption{Quantitative results on the COPD and COVID-19 test sets. IOU and ASSD (in mm) metrics are given in mean $\pm$ standard deviation. Boldface denotes the result significantly better than others ($p < 0.01$ with Bonferroni correction).} 
\centering  
\begin{minipage}[t]{0.99\linewidth}
    \centering
    (a) COPD results\\
    \vspace*{2mm}
{\resizebox{\textwidth}{!}{\begin{tabular}{l|l|l|l|l|l|l|l|l|l|l|l}
\toprule
Method&MAC&\#Param&Metric&Overall&Lungs&LUL&LLL&RUL&RLL& RML&interlobar\\
\midrule
  3DU-Net \cite{cciccek20163d}&10.5G&16.32M& \makecell{IOU \\ ASSD}&\makecell{0.915$\pm$0.037\\ 1.214$\pm$0.948}&\makecell{0.965$\pm$0.007 \\ 0.514$\pm$0.202}&\makecell{0.944$\pm$0.033\\0.766$\pm$0.839}&\makecell{0.937$\pm$0.007\\0.951$\pm$1.017}&\makecell{0.918$\pm$0.043\\1.264$\pm$1.050}&\makecell{0.937$\pm$0.032\\0.936$\pm$1.069}&\makecell{0.840$\pm$0.032\\2.153$\pm$2.738}&\makecell{N/A\\2.054$\pm$1.691} \\
\hline
  FRV-Net \cite{ferreira2018end}&~7.2G&15.5M&\makecell{IOU\\ASSD}&\makecell{0.918$\pm$0.038\\1.408$\pm$1.190}&\makecell{0.965$\pm$0.014\\0.602$\pm$0.393}&\makecell{0.950$\pm$0.038\\ 0.818$\pm$1.535}&\makecell{0.932$\pm$0.050\\1.030$\pm$1.481}&\makecell{0.917$\pm$0.050\\ 1.557$\pm$1.779}&\makecell{0.942$\pm$0.033\\0.957$\pm$1.414}&\makecell{0.848$\pm$0.103\\2.680$\pm$3.441}&\makecell{N/A\\2.292$\pm$2.218}\\
 \hline
  PDV-Net \cite{imran2019fast}&~7.2G&15.5M&\makecell{IOU\\ASSD}&\makecell{0.912$\pm$0.049\\ 3.027$\pm$5.544}&\makecell{0.951$\pm$0.031\\1.665$\pm$2.982}&\makecell{0.937$\pm$0.032\\ 1.802$\pm$2.926}&\makecell{0.926$\pm$0.044\\2.772$\pm$6.286}&\makecell{0.912$\pm$0.066\\2.885$\pm$6.600}&\makecell{0.926$\pm$0.050\\3.109$\pm$7.520}&\makecell{0.854$\pm$0.109\\4.540$\pm$7.800}&\makecell{N/A\\2.541$\pm$3.460}\\
   \hline
  RTSU-Net (ours)& 14.5G & 13.1M&\makecell{IOU\\ASSD}&\makecell{\textbf{0.949$\pm$0.026}\\ \textbf{0.607$\pm$0.537}}&\makecell{\textbf{0.976$\pm$0.010}\\ \textbf{0.326$\pm$0.166}}&\makecell{\textbf{0.962$\pm$0.020}\\ \textbf{0.482$\pm$0.534}}&\makecell{\textbf{0.959$\pm$0.023}\\ \textbf{0.465$\pm$0.446}}&\makecell{\textbf{0.952$\pm$0.030}\\ \textbf{0.668$\pm$1.020}}&\makecell{\textbf{0.960$\pm$0.010}\\ \textbf{0.534$\pm$0.518}}&\makecell{\textbf{0.912$\pm$0.080}\\ \textbf{0.885$\pm$1.412}}&\makecell{N/A\\ \textbf{0.947$\pm$0.800}}\\ 
\midrule
\bottomrule
\end{tabular}
}
}
\end{minipage}
\begin{minipage}[t]{0.99\linewidth}
    \centering
    \vspace*{2mm}
    (b) COVID-19 results \\
    \vspace*{2mm}
{\resizebox{\textwidth}{!}{\begin{tabular}{l|l|l|l|l|l|l|l|l|l|l|l}
\toprule
Method&MAC&\#Param&Metric&Overall&Lungs&LUL&LLL&RUL&RLL&RML&interlobar\\
\midrule
  3DU-Net \cite{cciccek20163d}&10.5G&16.3M&\makecell{IOU\\ASSD}&\makecell{0.904$\pm$0.051\\1.388$\pm$1.055}&\makecell{0.946$\pm$0.030\\0.840$\pm$0.666}&\makecell{0.936$\pm$0.031\\0.894$\pm$0.910}&\makecell{0.890$\pm$0.104\\1.491$\pm$1.560}&\makecell{0.911$\pm$0.068\\1.392$\pm$1.753}&\makecell{0.914$\pm$0.075\\1.454$\pm$2.215}&\makecell{0.870$\pm$0.093\\1.710$\pm$1.806}&\makecell{N/A\\2.213$\pm$2.025}\\
\hline
  FRV-Net \cite{ferreira2018end}&~9.3G&15.5M&\makecell{IOU\\ASSD}&\makecell{0.905$\pm$0.049\\1.236$\pm$1.058}&\makecell{0.952$\pm$0.029\\0.711$\pm$0.742}&\makecell{0.936$\pm$0.029\\0.796$\pm$0.695}&\makecell{0.907$\pm$0.075\\1.248$\pm$1.607}&\makecell{0.909$\pm$0.065\\1.346$\pm$1.569}&\makecell{0.914$\pm$0.083\\1.250$\pm$2.484}&\makecell{0.862$\pm$0.096\\1.541$\pm$1.500}&\makecell{N/A\\1.950$\pm$1.934}\\
\hline
  PDV-Net \cite{imran2019fast}&~9.3G&15.5M&\makecell{IOU\\ASSD}&\makecell{0.891$\pm$0.051\\1.908$\pm$1.727}&\makecell{0.943$\pm$0.030\\0.877$\pm$0.771}&\makecell{0.927$\pm$0.035\\1.379$\pm$2.158}&\makecell{0.885$\pm$0.086\\1.582$\pm$1.596}&\makecell{0.896$\pm$0.075\\3.451$\pm$5.194}&\makecell{0.903$\pm$0.082\\1.425$\pm$2.409}&\makecell{0.844$\pm$0.097\\1.705$\pm$1.464}&\makecell{N/A\\2.718$\pm$2.343}\\
\hline
  RTSU-Net (ours)& 14.5G & 13.1M&\makecell{IOU\\ASSD}&\makecell{\textbf{0.922$\pm$0.040}\\ \textbf{0.866$\pm$0.729}}&\makecell{\textbf{0.956$\pm$0.020}\\ \textbf{0.581$\pm$0.425}}&\makecell{\textbf{0.944$\pm$0.020}\\ \textbf{0.603$\pm$0.310}}&\makecell{\textbf{0.922$\pm$0.041}\\ \textbf{0.793$\pm$0.586}}&\makecell{\textbf{0.927$\pm$0.061}\\ \textbf{0.969$\pm$1.391}}&\makecell{\textbf{0.924$\pm$0.061}\\ \textbf{0.917$\pm$1.348}}&\makecell{\textbf{0.893$\pm$0.082}\\ \textbf{1.049$\pm$1.284}}&\makecell{\textbf{N/A}\\ \textbf{1.226$\pm$1.508}}\\ 
\midrule
\bottomrule
\multicolumn{12}{p{.9\textwidth}}{\rule{0pt}{2ex}  RUL, RML, RLL, LUL, LLL: Right upper, Right middle, Right lower, Left upper, Left lower lobes. Overall: per-lobe mean.}
\end{tabular}
}
}
\end{minipage}
\label{tab:QuantitativeResults}
\end{table*}

Table \ref{tab:QuantitativeResults} reports the quantitative results on both data sets.
The proposed method significantly outperformed the baseline methods and two published end-to-end lobe segmentation methods on both data set ($p < 0.01$ with Bonferroni correction) consistently in all measurements. Our model also exhibits  more robust performance, considering the smaller standard deviations. 

Box and whisker plots are provided in Fig.~\ref{fig:boxjaccard}. These plots show that for both the COVID-19 and the COPD cases, the right middle lobe is the most difficult to segment, which is not surprising given its known high variance in shape and the fact that the minor fissure is often incomplete or even absent. RTSU-Net clearly outperforms the other methods on both data sets. It can be also observed that there are less outliers with low IOU, indicating RTSU-Net is more  robust.

In terms of computational efficiency, the proposed method consumes even less memory than the baseline approach, with only a slight increase in the Multi-Adds operations (MAC). Hence, we conclude that the proposed method outperforms the other methods without introducing a substantial computational overhead. The proposed method processes a single scan at test time in 30 seconds on average, of which around 20 seconds are spent on model inference and the remainder on pre- and post-processing. 

\subsection{Ablation study}
Table \ref{tab:ablation} shows the results of the ablation study, where we compare the two-stage cascading framework without non-local modules, the framework with non-local modules without the geometric term, and the RTSU-Net. The results on both the COPD and COVID-19 data demonstrate the added value of the non-local module and show that the introduction of the geometric features increases the performance over the non-local module alone. This effect is most pronounced for the surface distance metric.

\begin{table}
\caption{Ablation study on the both data set for the non-local module (Non-local) and the geometric features (Geometric) into the two-stage cascading framework. Boldface denotes that a result is significantly better than others in the same column ($p < 0.01$ with Bonferroni correction).} 
\centering  
\begin{minipage}[t]{0.99\linewidth}
 \centering
 \vspace*{2mm}
    (a) COPD results\\
    \vspace*{2mm}
\begin{tabular}{m{6.5em}|m{1.5em}|m{1.5em}|m{2em}|m{5.5em}|m{4.5em}}
\toprule
Method & Two-stage & Non-local &Geo-metric & ASSD & IOU \\
\midrule
only two-stage  &\checkmark & & &1.122$\pm$1.315&0.940$\pm$0.031\\
\hline
w/o geometric &\checkmark &\checkmark &  &0.956$\pm$1.395 & 0.942$\pm$0.031\\
\hline
RTSU-Net &\checkmark &\checkmark &\checkmark &\textbf{0.607}$\pm$\textbf{0.537}& \textbf{0.949$\pm$0.026}\\
\midrule
\bottomrule
\end{tabular}
\end{minipage}
\begin{minipage}[t]{0.99\linewidth}
 \centering
 \vspace*{2mm}
    (b) COVID-19 results \\
 \vspace*{2mm}

\begin{tabular}{m{6.5em}|m{1.5em}|m{1.5em}|m{2em}|m{5.5em}|m{4.5em}}
\toprule
Method & Two-stage & Non-local &Geo-metric & ASSD & IOU \\
\midrule
only two-stage  &\checkmark & & &1.025$\pm$0.893&0.916$\pm$0.045\\
\hline
w/o geometric &\checkmark &\checkmark &  &1.370$\pm$1.582 & 0.918$\pm$0.037\\
\hline
RTSU-Net &\checkmark &\checkmark &\checkmark &\textbf{0.866}$\pm$\textbf{0.729}& \textbf{0.922$\pm$0.04}\\
\midrule
\bottomrule
\end{tabular}
\end{minipage}
\label{tab:ablation}
\end{table}

\subsection{Effect of the Non-Local module}
\begin{figure}[t]
\centering
\includegraphics[width=0.45\textwidth]{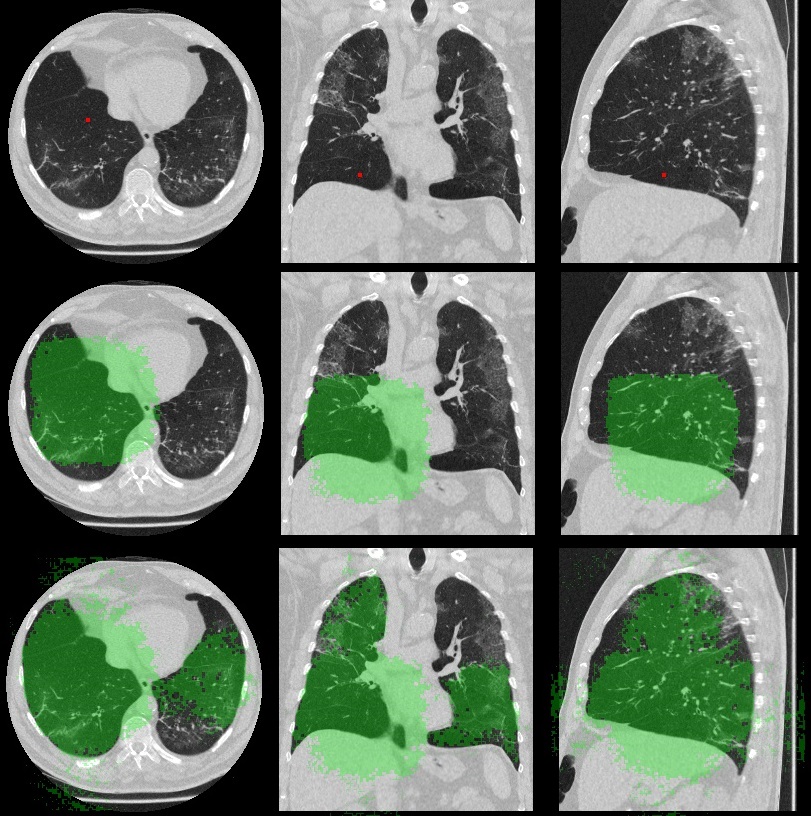}
\caption{Effective Receptive Field (ERF) before the non-local module (2nd row) and after (3rd row) by running forward pass for the first RU-Net on a CT scan from the COVID-19 test set. The green area indicates non-zero gradients (with respect to the input scan) of a feature at a location in the input scan corresponding to the red square (1st row).}
\label{fig:NonLocalFOV}
\end{figure}
\begin{figure}[t]
\centering
\begin{tabular}{c}
\includegraphics[width=0.45\textwidth]{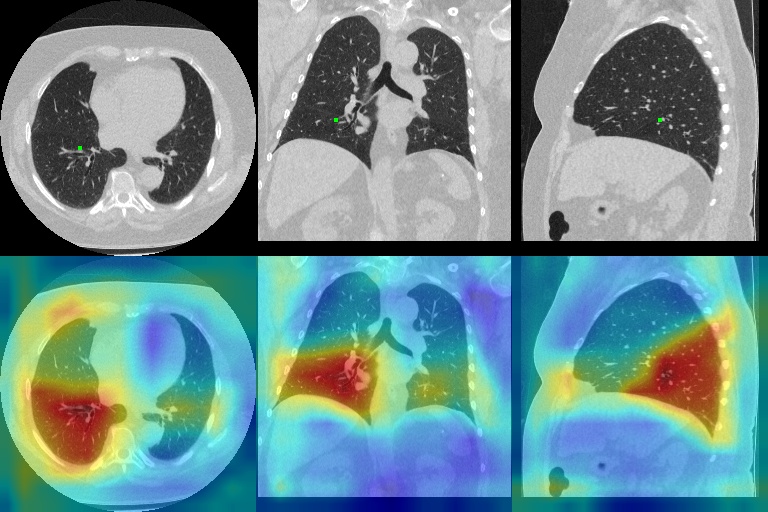} \\
(a) intra-lobe dependency \\
\includegraphics[width=0.45\textwidth]{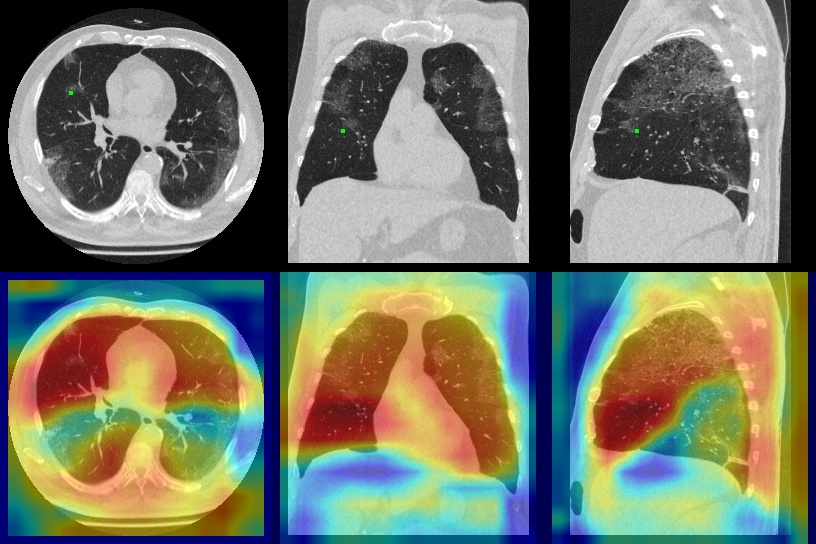} \\
(b) long-range dependency \\
\end{tabular}
\caption{The Self-Attention weights (2nd row) from the proposed non-local module for the feature whose location is shown using the green spot in the original input scan (1st row). We use color map jet \cite{zhou2015survey} for this plot. Two scans from the COVID-19 test set are shown. (a) demonstrates mostly the feature dependencies within the lobe in the clear lung. (b) indicates long-range dependencies are required when the target lobe is affected by disease.} \end{figure}\label{fig:self-attention}

In theory, the proposed non-local module can achieve a global receptive field in an efficient way instead of using aggressively down-sampled input or relying on much deeper CNN architectures. To measure the effective receptive field (ERF) size before and after the non-local operation, we computed the gradients $\frac{\partial F}{\partial I}\big|_{i}$ of the feature at the location $i$ in the feature map $F$ to the input image $I$. We run a forward pass for the first RU-Net on a CT scan from the COVID-19 test set. The ERF of the features at the same corresponding location before and after the non-local operation are visualized in Fig. \ref{fig:NonLocalFOV} for three orthogonal slices. 

The figure renders non-zero gradients in green and indicates the center of the ERF with a red square. The center is a mapped coordinate from the chosen feature in the feature map to the input image via up-sampling. Thus a slight shift may occur. The left image shows the ERF before the non-local operation is contained in a square due to the nature of stacked convolutions. However, the ERF after non-local on the right side shows a non-square distribution, reaching the other side of the lung. We, therefore, conclude that the non-local module can enlarge the effective receptive field dramatically.

To study the structured relationships between features, we visualize the self-attention weights for the feature at location $i$ given the feature map $x$ and geometric features $\mu$. We run a forward pass for the first stage RU-Net on two CT scans from the COVID-19 test set. The attention weights are the $i$-th row vector in the self-attention matrix corresponding to $f(x_{i}, x_{j}) + \tau(\mu_{i},\mu_{j})$ from the equation \ref{non-local-geo}. Figure \ref{fig:self-attention} (a) shows the feature at the location $i$ (green dot) mostly depends on the information within the lobe when the health lung is present. We can also clearly see the attention weights follows the lobe borders. Figure \ref{fig:self-attention} (b) shows a case with multiple ground-glass lesions, where the interactions between the  feature representing the region nearby the right middle lobe and features presenting other regions in the entire lung. Interestingly, we note that by introducing the geometric term in the non-local module, attention weights also correspond to the lung bounding box. 
\begin{figure*}[ht!]
\centering
\setlength{\tabcolsep}{0.001\textwidth}
\begin{tabular}{cccccc}

  \includegraphics[width=0.165\textwidth, height=3.5cm]{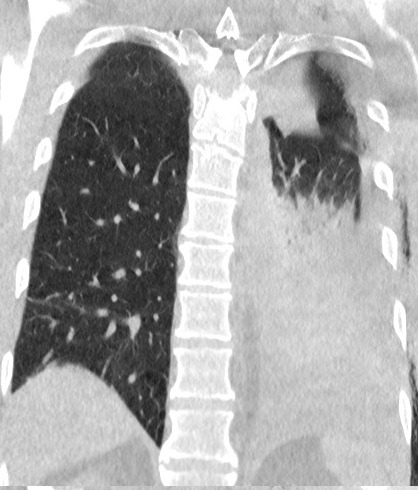}
  &\includegraphics[width=0.165\textwidth, height=3.5cm]{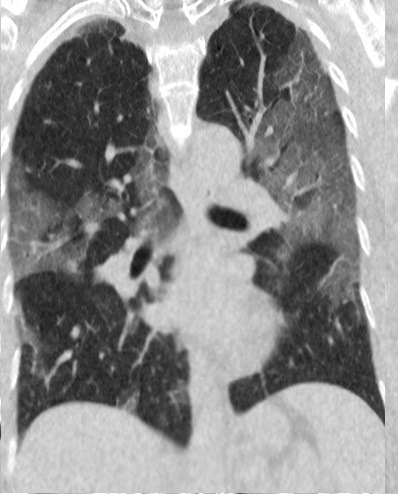}
  &\includegraphics[width=0.165\textwidth, height=3.5cm]{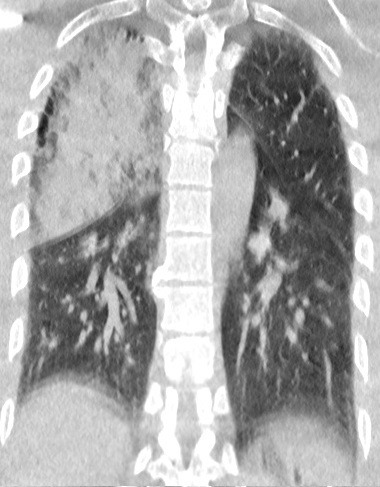}
  &\includegraphics[width=0.165\textwidth, height=3.5cm]{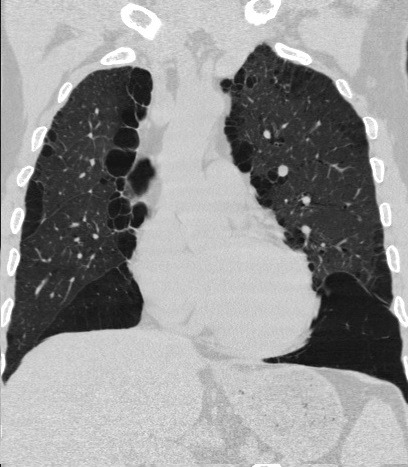}
  &\includegraphics[width=0.165\textwidth, height=3.5cm]{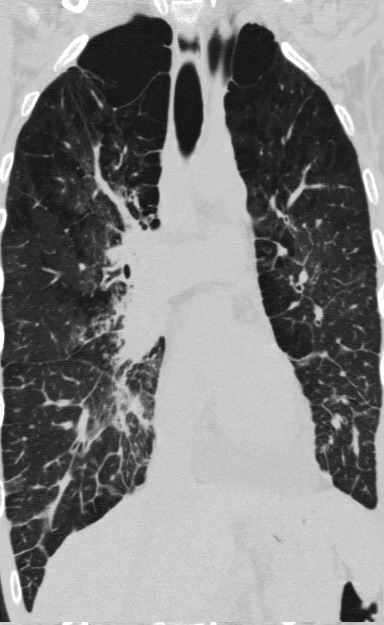} 
  &\includegraphics[width=0.165\textwidth, height=3.5cm]{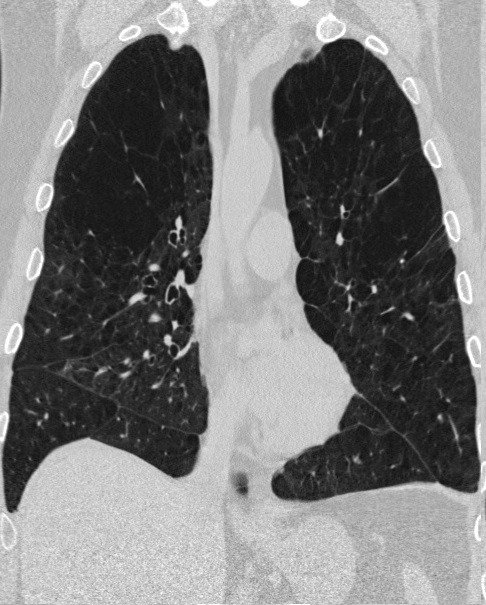}\\
  \includegraphics[width=0.165\textwidth, height=3.5cm]{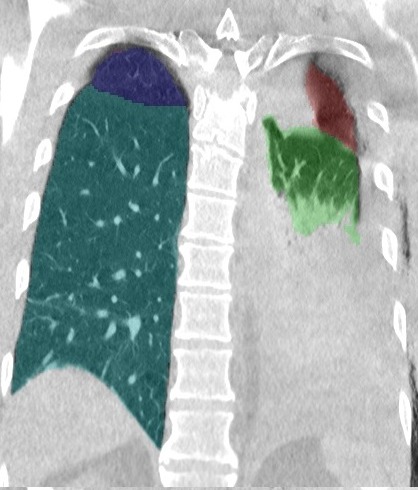}
  &\includegraphics[width=0.165\textwidth, height=3.5cm]{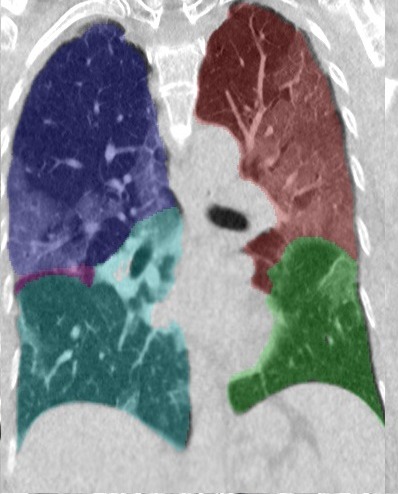}
  &\includegraphics[width=0.165\textwidth, height=3.5cm]{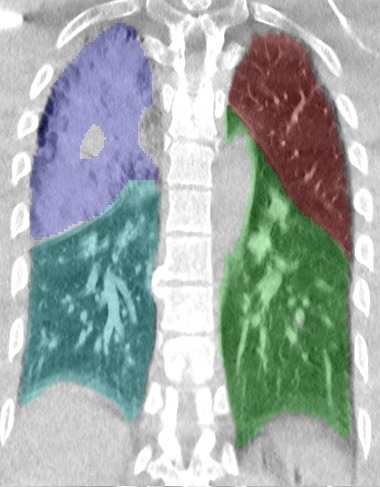}
  &\includegraphics[width=0.165\textwidth, height=3.5cm]{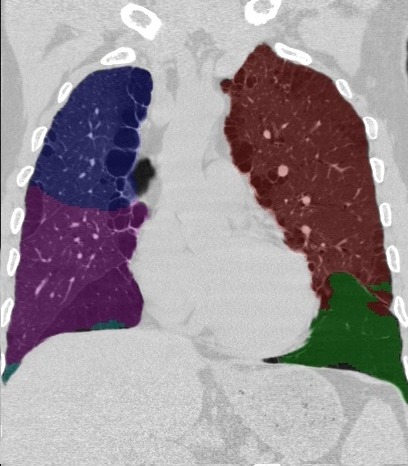} 
  &\includegraphics[width=0.165\textwidth, height=3.5cm]{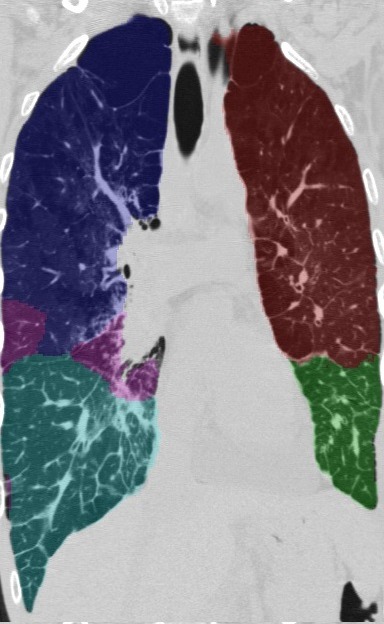}
  &\includegraphics[width=0.165\textwidth, height=3.5cm]{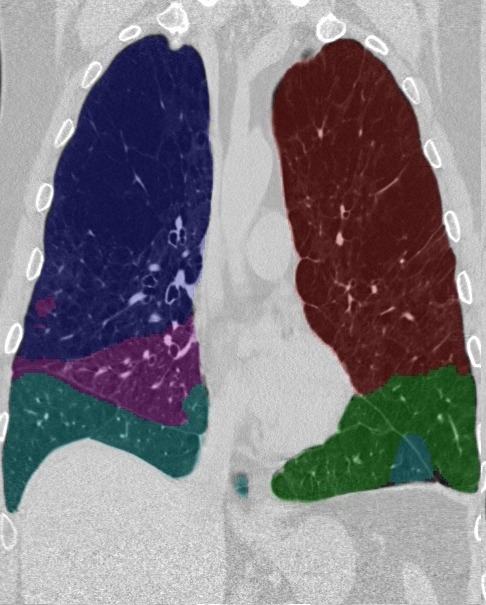}\\
  \includegraphics[width=0.165\textwidth, height=3.5cm]{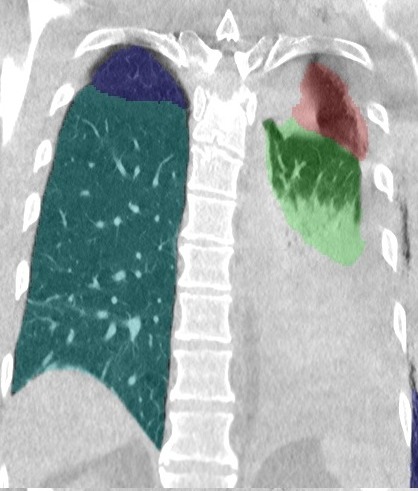}
  &\includegraphics[width=0.165\textwidth, height=3.5cm]{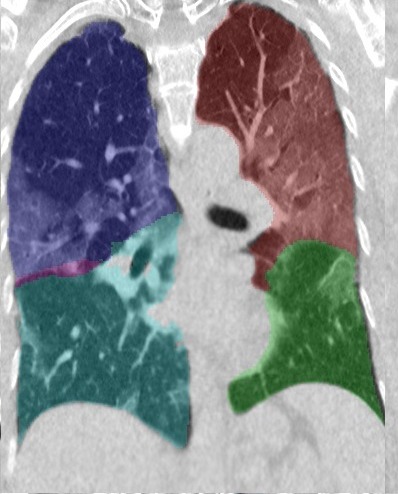}
  &\includegraphics[width=0.165\textwidth, height=3.5cm]{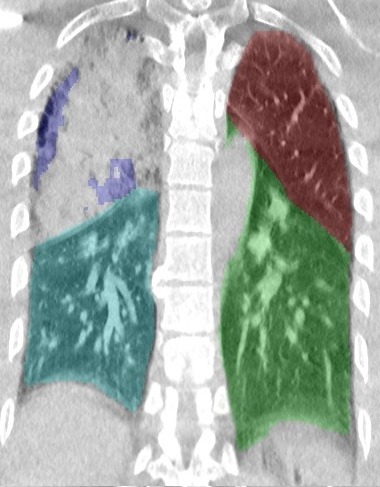}
  &\includegraphics[width=0.165\textwidth, height=3.5cm]{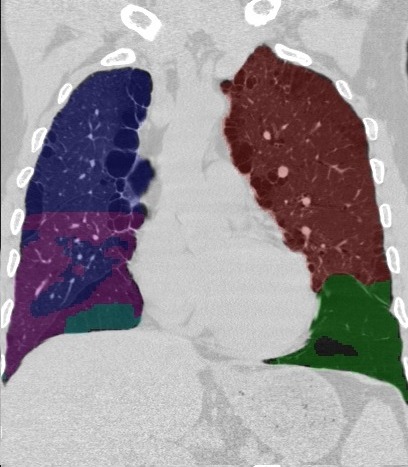}
  &\includegraphics[width=0.165\textwidth, height=3.5cm]{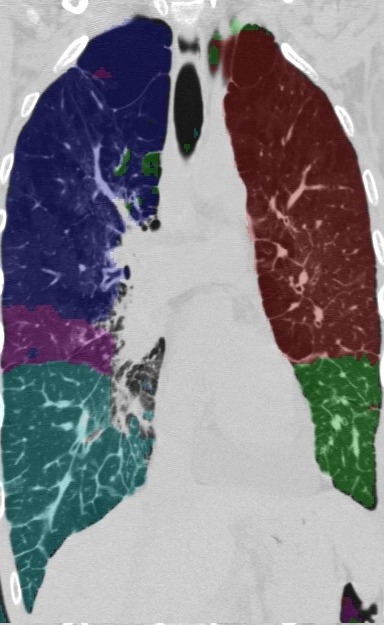}
  &\includegraphics[width=0.165\textwidth, height=3.5cm]{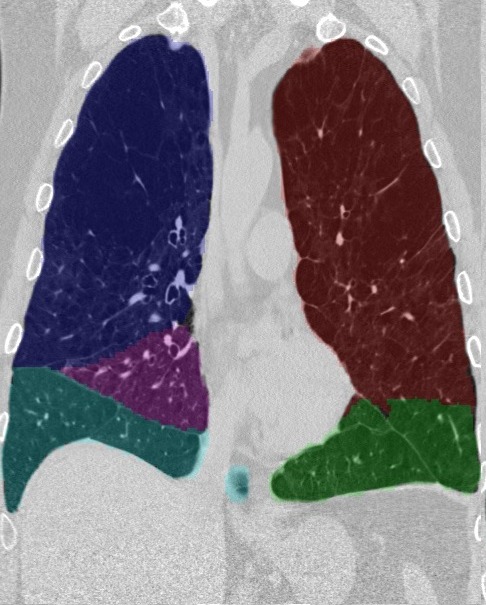}\\
  \includegraphics[width=0.165\textwidth, height=3.5cm]{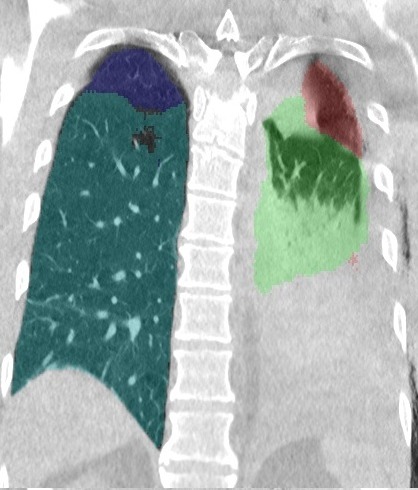}
  &\includegraphics[width=0.165\textwidth, height=3.5cm]{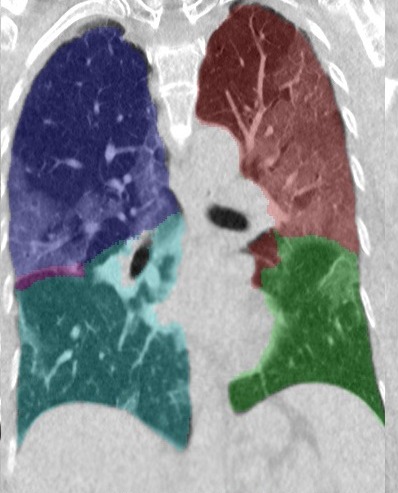} 
  &\includegraphics[width=0.165\textwidth, height=3.5cm]{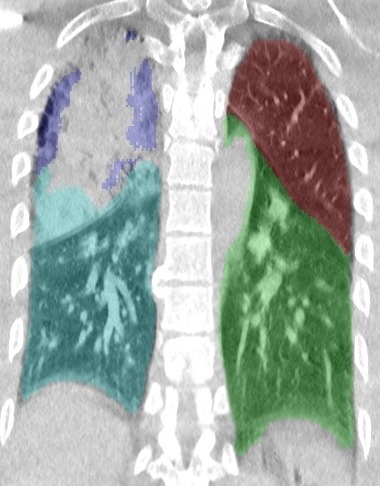}
  &\includegraphics[width=0.165\textwidth, height=3.5cm]{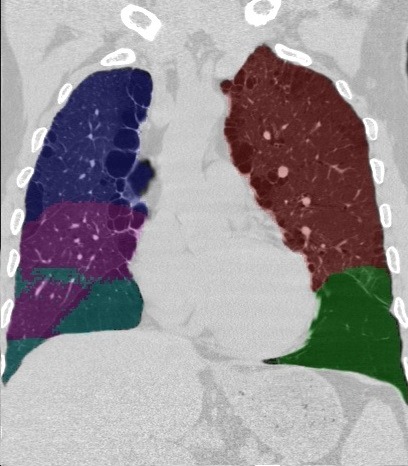}
  &\includegraphics[width=0.165\textwidth, height=3.5cm]{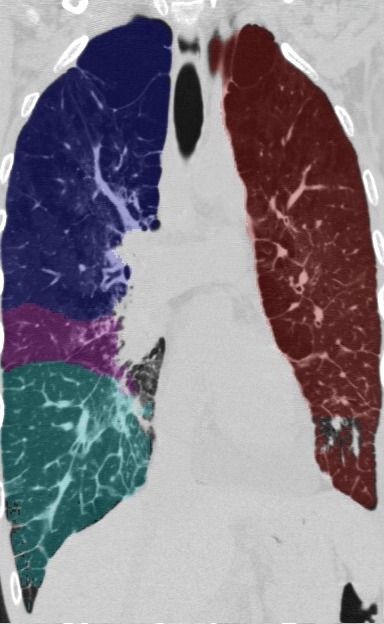}
  &\includegraphics[width=0.165\textwidth, height=3.5cm]{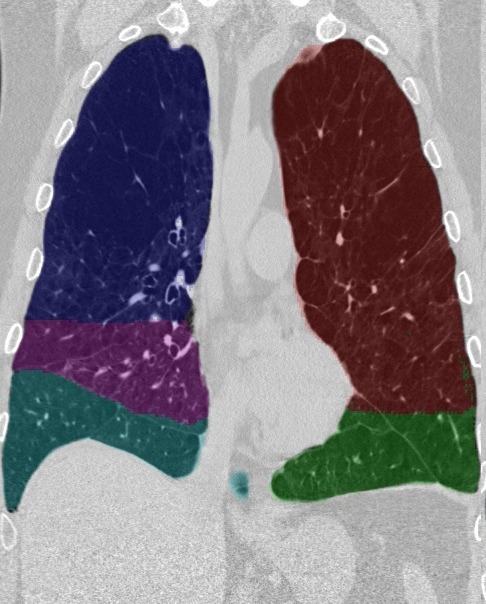}\\
  \includegraphics[width=0.165\textwidth, height=3.5cm]{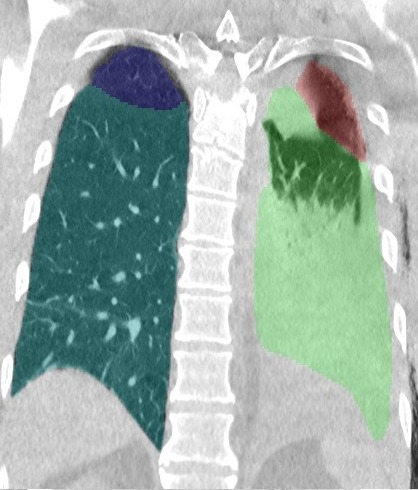} 
  &\includegraphics[width=0.165\textwidth, height=3.5cm]{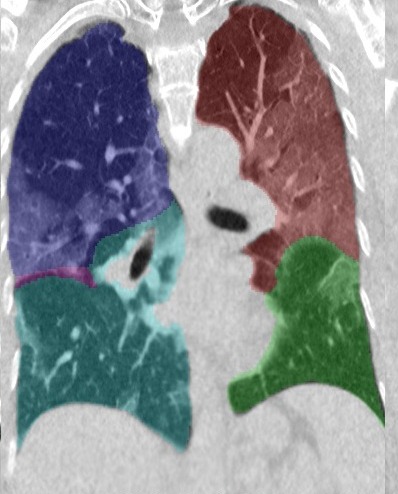} 
  &\includegraphics[width=0.165\textwidth, height=3.5cm]{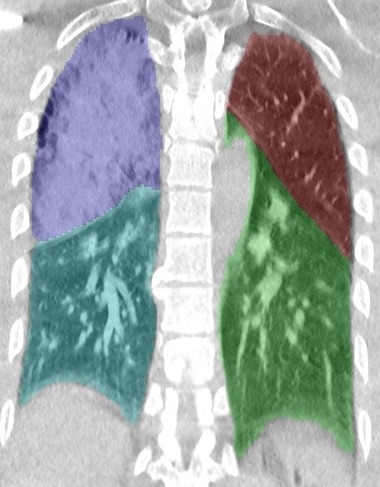} 
  &\includegraphics[width=0.165\textwidth, height=3.5cm]{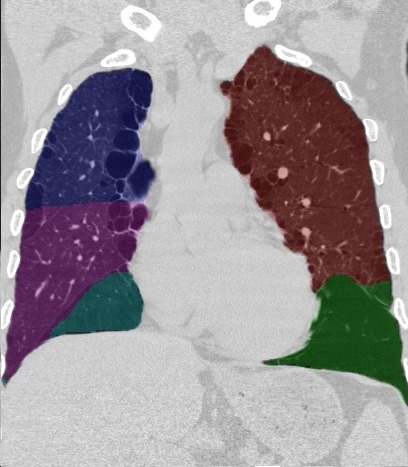} 
  &\includegraphics[width=0.165\textwidth, height=3.5cm]{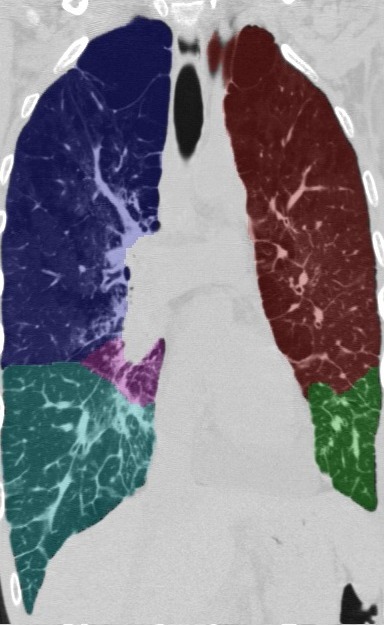} 
  &\includegraphics[width=0.165\textwidth, height=3.5cm]{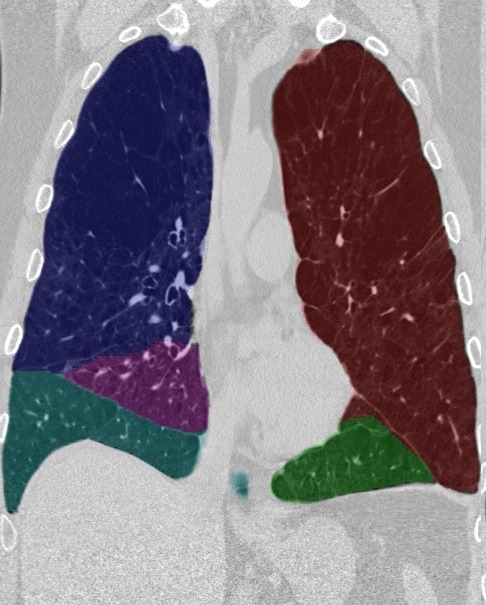}\\
   \includegraphics[width=0.165\textwidth, height=3.5cm]{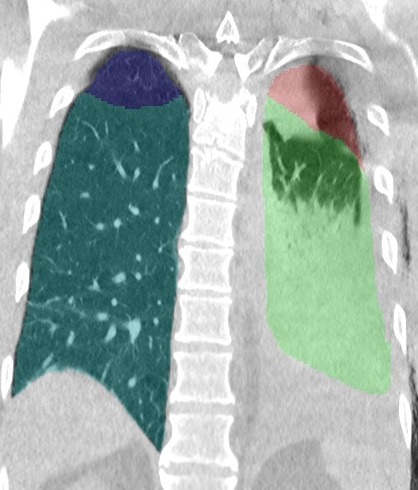} 
  &\includegraphics[width=0.165\textwidth, height=3.5cm]{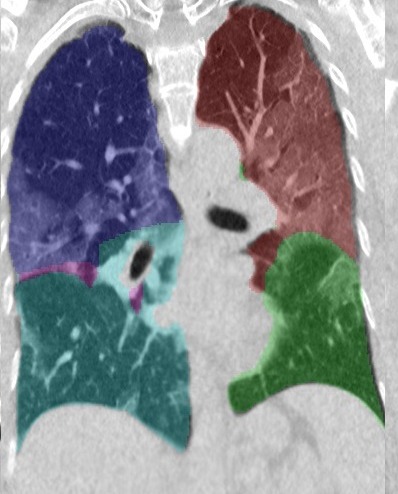} 
  &\includegraphics[width=0.165\textwidth, height=3.5cm]{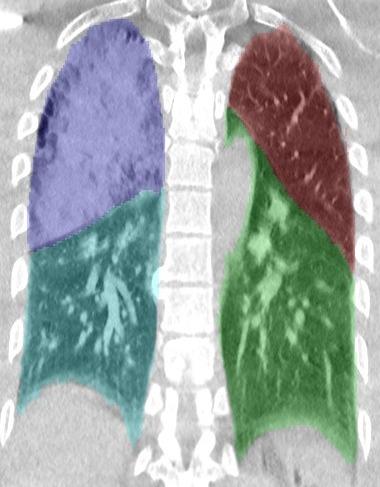} 
  &\includegraphics[width=0.165\textwidth, height=3.5cm]{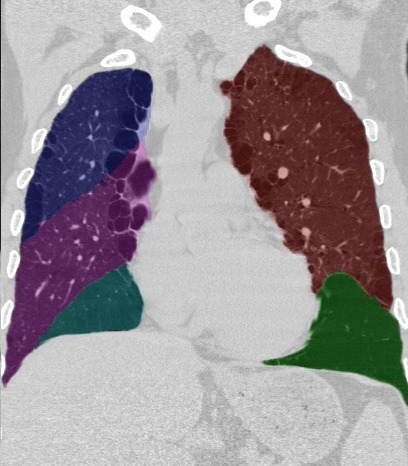} 
  &\includegraphics[width=0.165\textwidth, height=3.5cm]{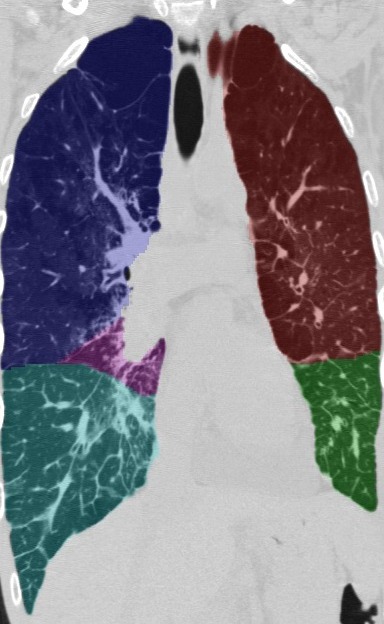} 
  &\includegraphics[width=0.165\textwidth, height=3.5cm]{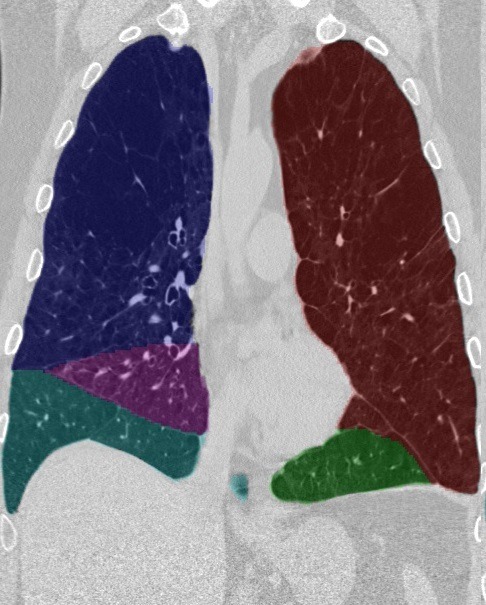}\\
\end{tabular}
\caption{Qualitative comparison of segmentation results for six representative test cases. The left three columns show COVID-19 cases, the right three columns show COPD cases. From top to bottom: input image, 3DU-Net baseline, PDV-Net, FRV-Net, the proposed RTSU-Net, and the segmentation reference. \textcolor{RULC}{\rule{.2cm}{.2cm}} right upper, \textcolor{RMLC}{\rule{.2cm}{.2cm}} right middle, \textcolor{RLLC}{\rule{.2cm}{.2cm}} right lower,
\textcolor{LULC}{\rule{.2cm}{.2cm}} left upper, \textcolor{LLLC}{\rule{.2cm}{.2cm}} left lower lobes.} \label{fig:qresult}
\end{figure*}
\subsection{Qualitative Results}
Fig.~\ref{fig:qresult} shows results for the 3D U-Net, PDV-Net \cite{imran2019fast}, FRV-Net \cite{ferreira2018end}, and the RTSU-Net from top to bottom. For comparison, reference segmentations are provided in the bottom row. We selected three COPD (4-6 column) and three COVID-19 cases (1-3 column) with various levels of pathological and anatomical variations. We observed that all methods usually do not produce oversegmentation of the lungs. By capturing feature dependencies, we see that the proposed method generates generally smoother lobe borders and is even able to infer the approximately correct lobe shapes when the lung is filled with fluid (1st column).

\subsection{Comparison with human readers}
\label{humancomparison}
To evaluate human performance, we asked two independent human readers (analysts) to manually segment the lobes from scratch, given segmentation of the lung. Their results are evaluated on a random set of 100 scans from the COPD test set. The human readers achieved $0.953\pm0.017$ IOU and $0.501\pm0.193$ ASSD (in mm) on average, while the RTSU-Net achieved $0.953\pm0.015$ IOU and $0.541\pm0.231$ ASSD. The human readers and the RTSU-Net method are both significantly better than the other methods. 

In terms of lung segmentation, the analysts reached $0.974\pm0.015$ IOU and $0.34\pm0.214$ ASSD on average, while the RTSU-Net achieved $0.977\pm0.009$ IOU and $0.325\pm0.2$ ASSD on average. 

Regarding the fissure alignment, the analysts reached $0.686\pm0.361$ ASSD on average while the RTSU-Net achieved $0.835\pm0.398$ ASSD on average. We conclude that the RTSU-Net method performs comparably to humans for segmenting the lung and the lobes.

\subsection{Validation on LOLA11}
We have applied our method to the 55 scans of the LOLA11 challenge, available on \texttt{https://lola11.grand-challenge.org/}. This is an independent test set in which approximately half of these scans are very difficult to segment due to the presence of gross pathology. Lobar borders are completely invisible in some of these scans. 

Our method (submission date May 3, 2020) reaches a mean IOU of 0.9197 for the lobe segmentation and 0.9706 in lung segmentation. This score is comparable to the other top participants and ranks \#2 for automatic lobe segmentation methods, after submission of a not yet published variant of LobeNet (submission date November 20, 2019). 

\section{Discussion and Conclusion}
We have presented a novel method using relational two-stage convolution neural networks for segmenting pulmonary lobes in CT images. The proposed method is capable of capturing visual and geometric correspondence between high-level convolution features, which may represent the relationships between objects and object parts. This proposed non-local module can also be used to effectively and efficiently enlarge the receptive field of convolution features. This module can be easily used as a common neural network layer in other computer vision tasks such as object detection and classification. 

We show in our results that learning feature dependencies improves the lobe segmentation performance significantly on the COPD and the COVID-19 data set. The average symmetric distance metric in the ablation study shows that using geometric features is effective for generating more precise object boundaries. This can also be observed from the qualitative results, where the lobe boundaries from the proposed method are more consistent with the reference lobe shapes. Without depending on prior lung segmentation, our approach serves as an end-to-end lobe segmentation framework that can be used for lung segmentation as well, by taking the union of lobes per lung.

In terms of computational efficiency, our method maintains the same level of Multi-Adds operations (MAC) as the standard 3D U-Net and two other approaches previously proposed for pulmonary lobe segmentation. It requires even fewer trainable parameters compared to the standard 3D U-Net. Our method can be trained and tested on a consumer-level GPU with 12 GB memory, and speed at test time is around 30 seconds for a full resolution CT scan (20 seconds for deep learning inference and 10 for pre-processing and post-processing). 

For segmenting the lobes in scans of COPD patients, the previously published LobeNet method \cite{gerard2019pulmonary} reported excellent performance on 1076 scans from COPDGene, with an ASSD of 0.138 mm, well below the voxel resolution and well below what RTSU-Net and independent human analysts achieved in a set of 100 COPDGene scans in this study (Sect.~\ref{humancomparison}). These metrics are not directly comparable as \cite{gerard2019pulmonary} used a different set of scans and a reference partly provided by a software package. For future studies, it would be interesting to directly compare both approaches. On LOLA11, LobeNet outperformed RTSU-Net by a very small margin. We noticed failures of RTSU-Net on scans with abnormalities distinct from what occurred in the COPD and COVID-19 training data. 

Segmentation of lobes in scans of patients with severe pneumonia due to COVID-19 is not an easy task. In this work, we used only 370 COVID-19 CT scans for training. Thanks to pre-training on 4000 COPD scans, we still obtained good results with a small training set, and we were able to provide lobe segmentations robust to the presence of ground-glass, consolidations, and crazy paving. 

Lobe segmentation is an important prerequisite for accurate quantification of lung damage in COVID-19 CT scans. Fig.~\ref{fig:qresult} shows that the standard 3D U-Net (2nd row), PDV-Net (3rd row), and FRV-Net (4th row) may miss areas of consolidation (3rd column) while the RTSU-Net found the lobes accurately. RTSU-Net also performs reasonably well when this lobe is completely filled with pleural fluid (first column). Nevertheless, we also see that sometimes the border of the segmentations of the proposed method is incorrect (3rd column, the right upper lobe shows a slight oversegmentation across the lobar borders towards the shoulder), 

We hypothesize that a larger training set would further improve performance, especially for cases with gross pathological changes that are not yet well represented in the current training scans. Nevertheless, the results presented here are sufficient for further analysis, and we believe that they will prove useful in automated per-lobe severity scoring. This is a topic for future research. 

We freely share our segmentation algorithm on \texttt{\small  https://grand-challenge.org/algorithms/} and provide results for public data such as the scans from \texttt{\small https://coronacases.org/}.

%


\ifCLASSOPTIONcaptionsoff
  \newpage
\fi



\bibliographystyle{IEEEtran}
\bibliography{Reference.bib}
%

\end{document}